# Chemisorption Induced Formation of Biphenylene Dimer on Surfaces

Zhiwen Zeng[1,‡], Dezhou Guo[2,‡], Tao Wang[1,3,4,*], Qifan Chen[5], Adam Matěj[5], Jianmin Huang[1], Dong Han[1], Qian Xu[1], Aidi Zhao[6], Pavel Jelínek[5], Dimas G. de Oteyza[3,4,7], Jean-Sabin McEwen[2,8,9,10,11,*], Junfa Zhu[1,*]

[1] National Synchrotron Radiation Laboratory, Department of Chemical Physics and Key Laboratory of Surface and Interface Chemistry and Energy Catalysis of Anhui Higher Education Institutes, University of Science and Technology of China, Hefei, 230029, P. R. China.

[2] The Gene & Linda Voiland School of Chemical Engineering and Bioengineering, Washington State University, Pullman, WA 99164, United States.

[3] Donostia International Physics Center, San Sebastián 20018, Spain.

[4] Centro de Fisica de Materiales, CFM/MPC, CSIC-UPV/EHU, San Sebastián 20018, Spain.

[5] Institute of Physics of the Czech Academy of Sciences, Cukrovarnická 10, 16200 Prague 6, Czech.

[6] School of Physical Science and Technology, ShanghaiTech University, Shanghai 201210, China.

[7] Ikerbasque, Basque Foundation for Science, 48013 Bilbao, Spain.

[8] Institute for Integrated Catalysis, Pacific Northwest National Laboratory, Richland, WA 99352, United States.

[9] Department of Physics and Astronomy, Washington State University, Pullman, WA 99164, United States.

[10] Department of Chemistry, Washington State University, Pullman, WA 99164, United States.

[11] Department of Biological Systems Engineering, Washington State University, Pullman, WA 99164, United States.

*KEYWORDS on-surface synthesis, adsorption, biphenylene dimer, anti-aromaticity, reaction selectivity, scanning tunneling microscopy, synchrotron radiation photoemission spectroscopy*

**ABSTRACT:** We report an example that demonstrates the clear interdependence between surface-supported reactions and molecular adsorption configurations. Two biphenyl-based molecules with two and four bromine substituents, *i.e.* 2,2'-dibromo-biphenyl (DBBP) and 2,2',6,6'-tetrabromo-1,1'-biphenyl (TBBP), show completely different reaction pathways on a Ag(111) surface, leading to the selective formation of dibenzo[e,l]pyrene and biphenylene dimer, respectively. By combining low-temperature scanning tunneling microscopy, synchrotron radiation photoemission spectroscopy, and density functional theory calculations, we unravel the underlying reaction mechanism. After debromination, a bi-radical biphenyl can be stabilized by surface Ag adatoms, while a four-radical biphenyl undergoes spontaneous intramolecular annulation due to its extreme instability on Ag(111). Such different chemisorption-induced precursor states between DBBP and TBBP consequently lead to different reaction pathways after further annealing. In addition, using bond-resolving scanning tunneling microscopy and scanning tunneling spectroscopy, we determine the bond length alternation of biphenylene dimer product with atomic precision, which contains four-, six-, and eight-membered rings. The four-membered ring units turn out to be radialene structures.

## Introduction

On-surface synthesis (OSS) has shown its great potential in the fabrication of functional molecules and covalent nanostructures with atomic precision in the last decade.[1-3] Different from solution phase chemistry, due to the required clean reaction environment (ultrahigh vacuum) at the gas-solid interface, the use of catalysts is largely limited in OSS. Hence, steering reaction pathways in OSS has been more challenging overall than that in wet chemistry. Chemical organic reactions on surfaces typically include three basic steps: molecular adsorption, diffusion and reaction. The reported examples toward steering reaction pathways on surfaces were mostly focused on tuning the molecular diffusion and the reaction barriers.[1-3] For instance, it has been demonstrated that self-assembly templates can efficiently direct the reaction pathways by confining molecular diffusion.[4-7] In addition, different metal substrates or metal adatoms normally have a different catalytic activity toward a specific on-surface reaction, thus impacting the reaction barriers.[8-11] However, related studies on the adsorption process are rare, although it actually differentiates the heterogeneous on-surface synthesis from the homogenous solution phase chemistry. A few reported examples were focused on the physisorption of intact molecules on surfaces. The adsorption height from the surface[12] as well as the adsorption site[13] can play important roles on the reactivity of the functional groups of precursor molecules.

**Scheme 1. Reaction pathways of (a) DBBP and (b) TBBP on the Ag(111) surfaces, respectively.**

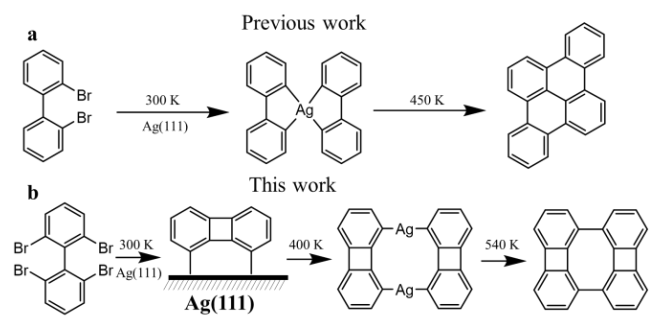

The coupling reactions typically involve the generation and coupling of radicals and have been extensively studied.[2, 14-15] Radicals are generated when functional groups are activated. The adsorption configuration of newly formed radical species is very different from that of the initial molecule because active radical species are normally stabilized by surface atoms or stray surface adatoms.[16] Thus, the adsorption configurations of the activated molecules are largely determined by their radical sites. As a result, it is reasonable to infer that molecule with the same backbone, but different number of radicals may lead to a dramatically different adsorption behavior on surfaces after activation. In turn, the different adsorption behaviors may potentially influence the reaction pathways and the final products.

Herein, we report such an example by comparing the reactions of 2,2'-dibromo-biphenyl (DBBP) and 2,2',6,6'-tetrabromo-1,1'-biphenyl (TBBP) molecules on a Ag(111) surface. In our previous work,[17] we showed that the bi-radical biphenyl species formed upon debromination of DBBP were efficiently stabilized by surface Ag adatoms at 300 K. Further annealing led to the formation of dibenzo[e,l]pyrene nanographene. However, in this work, TBBP shows a unique reaction pathway: the first step involves the generation of four-radical biphenyl species after the debromination of the precursor; then TBBP undergoes intramolecular annulation spontaneously at 300 K due to its extreme instability on Ag(111), forming a bi-radical biphenylene monomer that is anchored to the surface; further annealing leads to the formation of an organometallic intermediate state, followed by its transformation into covalent biphenylene dimers containing 4, 6, and 8- membered carbon rings. The chemical structure and electronic properties of the biphenylene dimer have been studied by bond-resolving scanning tunneling microscopy (BR-STM) and scanning tunneling spectroscopy (STS), offering significant insights into its potential anti-aromaticity. The mechanism for the different reaction selectivity between DBBP and TBBP, *i.e.* the formation of dibenzo[e,l]pyrene *vs.* a biphenylene dimer, has been further studied by density functional theory (DFT) calculations. This work reveals that the chemisorption behavior of adsorbates can play a decisive role on the reaction pathway. In addition, the bond alternation of the intriguing biphenylene dimer,[18] as proposed by organic and theoretical chemists, has been corroborated here in real space. In fact, the fabrication of four-membered ring containing structures on surfaces has become a hot topic and been widely studied recently due to their exotic electronic and mechanical properties, which can be achieved by either intramolecular or intermolecular [2+2] annulation reactions.[19-26] An outstanding example was reported Fan *et al..*[24] In their work, the biphenylene network with periodically arranged four-, six-, and eight-membered rings was synthesized by inter-polymer hydrogen-fluoride zipping reaction and exhibits metallic electronic properties. The chemisorption induced formation of four-membered ring as presented in our work provides a new insight into the fabrication of four-membered ring containing functional nanostructures on surfaces.

**Results and Discussion**

**Synthesis of biphenylene dimer.** Figure 1 presents the experimental result of TBBP molecules adsorbing on Ag(111) at room temperature. TBBP molecules stay intact on Ag(111) at ~250 K, as revealed by Br 3$d$ and C 1$s$ synchrotron radiation photoemission (SRPE) spectra in Figure 1a and 1b. The ratio between C-Br and C-C is 1:2.4, in fair agreement with the ideal value of 1:2 as derived from the structural model shown in Figure 1c. The corresponding STM images are shown in Figure S1. The molecules self-assemble into square ordered islands. A single molecule is composed of one bright head (yellow dotted contour) and one weak tail (green dotted contour), indicating its twisted adsorption configuration because of the repulsion between the adjacent Br atoms of intact TBBP.[15, 27]

After depositing TBBP molecules onto the Ag(111) surface held at 300 K, the molecules exhibit bright protrusions in the STM images as shown in Figure 1d and 1e (an overview STM image is shown in Figure S2). Most molecules aggregate into close-packed islands (Figure 1d), together with a few sparsely distributed trimers and tetramers (Figure 1e). In particular, the existence of the trimer (Figure 1e) implies that the bright protrusion cannot from a submolecular feature, *i.e.* one protrusion cannot correspond to one phenyl group of TBBP. In addition, the center-to-center distance between adjacent bright protrusions is measured to be 7.8 Å, which is much larger than the distance between the two phenyls of TBBP (~4.3 Å). Therefore, one bright protrusion can only correspond to one individual TBBP molecule. The majority of C-Br bonds of TBBP are dissociated at 300 K, as evidenced by the downward shift of the Br 3$d$ core level binding energies from 250 K to 300 K.[28] The partial dissociation of C-Br bonds on Ag(111) was also reported by previous works.[2, 29-30] The Br adatoms on the surface can be recognized as the relatively dark and small dots in the STM images,[6, 31] as pointed out by the blue arrows in Figure 1e. Interestingly, the appearance of a new C 1$s$ component at a low binding energy of 283.0 eV at 300 K (Figure 1b) implies that radicals are probably stabilized by the surface atoms *via* a C-Ag coordination.[28, 31] The ratio between C-Ag and C-C is about 1:6 from the C 1s spectrum (C-Br: C-C: C-Ag=0.3: 6.5: 1), in agreement with the idea value in Figure 1c, implying that only two radicals are coordinated to the surface atoms. It is worth noting that the C 1s binding energy shift toward the low energy direction at 300 K with respect to that at 250 K is attributed to the increase of the surface work function induced by chemisorbed Br adatoms on the surface.[28-29, 32-34] Inspired by these findings and related previous works,[35] one can intuitively deduce that the four-radical biphenyl undergoes intramolecular annulation reactions at these conditions, forming biphenylene, while the two residual unquenched radicals are stabilized by the Ag surface. Consequently, it adsorbs perpendicularly to the surface, as schematically shown in Scheme 1. This is the reason why adsorbates show such bright features as compared to that of conventional flat molecules that adsorb parallel to the surface. A comparison between the apparent height of the bi-radical biphenylene monomer and the final biphenylene dimer product on Ag(111) is presented in Figure S3, where a difference of 1.8 Å is obtained. The perpendicular adsorption configuration also explains the formation of trimers and tetramers in Figure 1d and 1e, which should be stabilized by π-π stacking between face-to-face phenyls.[36-38] The DFT-calculated structural model of trimer and tetramer are displayed in Figure 1f and 1g. The distance between two adjacent molecules is about 7.8 Å for both trimer and tetramer, which is in excellent agreement with the experimental result (Figure 1e). The formation of biphenylene dimer is verified by our DFT calculations. Once TBBP molecules are fully debrominated, four-radical biphenyl species are not stable on the Ag surface but immediately start an intramolecular annulation reaction (Figure 1i and 1j). As a result, a biphenylene complex is formed with bi-radical binding with two surface silver atoms, leading to the non-radical side of the benzene ring pointing away from the surface. We tested various adsorption sites of the Ag (111) surface with biphenylene (as shown in Figure S8 and S9) and found that two radicals prefer to bind with two nearby surface Ag atoms rather than Ag adatom.

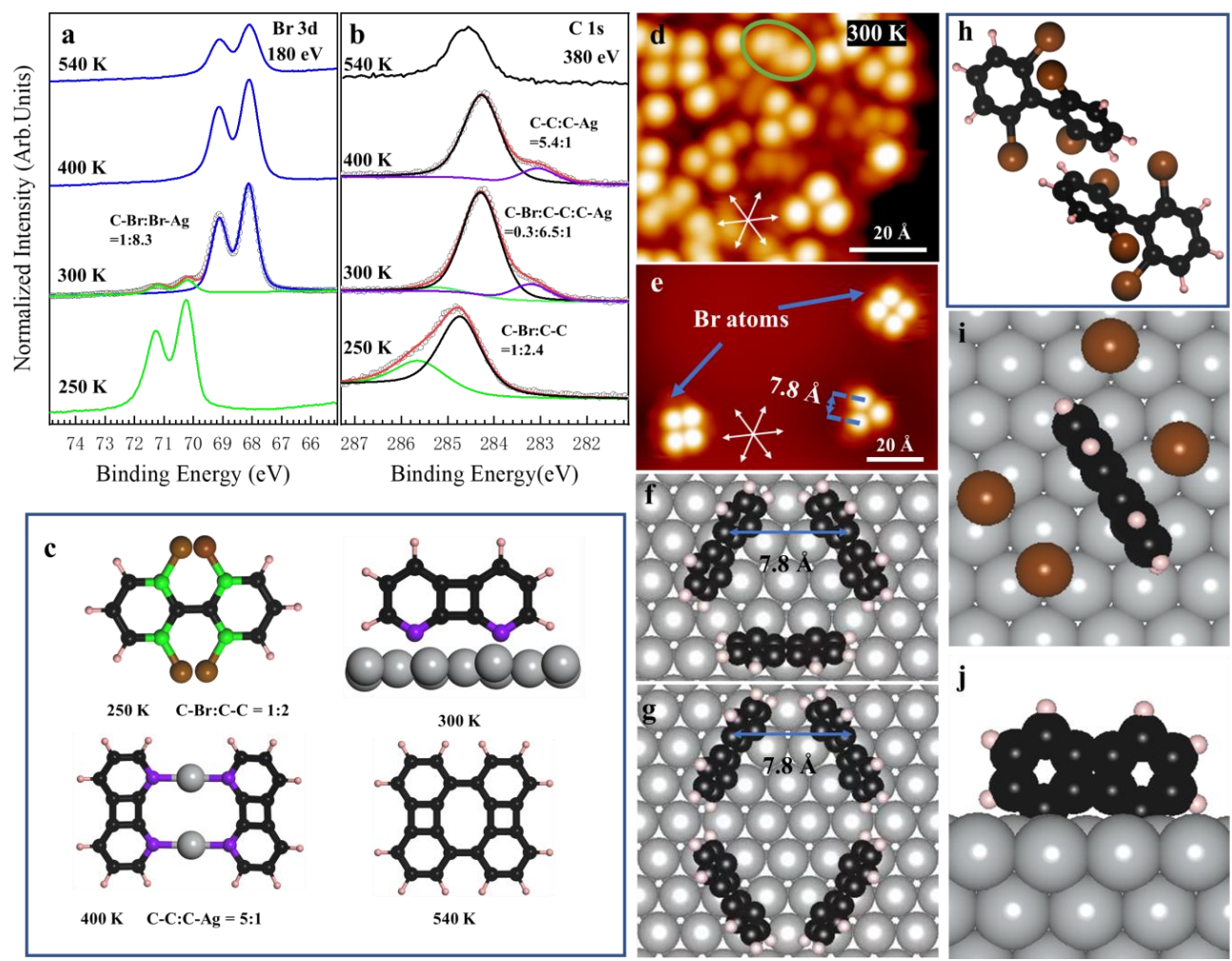

**Figure 1.** (a, b) Br 3d and C 1s SRPE spectra of the samples prepared by depositing TBBP on Ag(111) held at 250 K, held at 300 K followed by annealing to 400 K, 540 K. The photon energies for Br 3d and C 1s are 180 and 380 eV, respectively. (c) The molecular models of the major products in each temperature points. Different carbon atoms are depicted by different colors to illustrate their chemical environments. The ideal ratios of these C atoms are shown below each molecular model. (d, e) STM images of the sample upon deposition of TBBP on Ag(111) held at 300 K. The three high symmetric directions of the Ag(111) surface are shown as the white arrows; same for the following overview STM images. Tunneling parameters: (d, e) U = −1.5 V, I = 50 pA. (f, g) The DFT-calculated structural model of timer and tetramer in (e). (h) The possible structural model of two TBBP molecules in the green circle of (d). (i, j) Top and side views of DFT optimized structure of debrominated TBBP adsorbing on Ag(111). Color code: C, black; Ag, grey; H, pink; Br, brown; C bonded to Br atom, green; C bonded to Ag atom, purple.

Note that there are a few C-Br bonds remaining intact at 300 K from the XPS analysis, which could belong to the molecules inside the green circle in Figure 1d because of their similar STM morphology as the intact molecules at 250 K (Figure S1). Although the possibility that one or more Br atoms are lost on these molecules cannot be excluded, it is most probable that these molecules are attributed to intact TBBP. Because the π-π stacking interactions between TBBP and the neighboring biradical biphenylenes (Figure 1d) is different from the π-π interactions in the ordered TBBP island at 250 K (Figure S1), the morphology of TBBP in STM images can be slightly different.[39-40] Figure 1h shows the corresponding structural model of the two molecules in the green circle in Figure 1d.

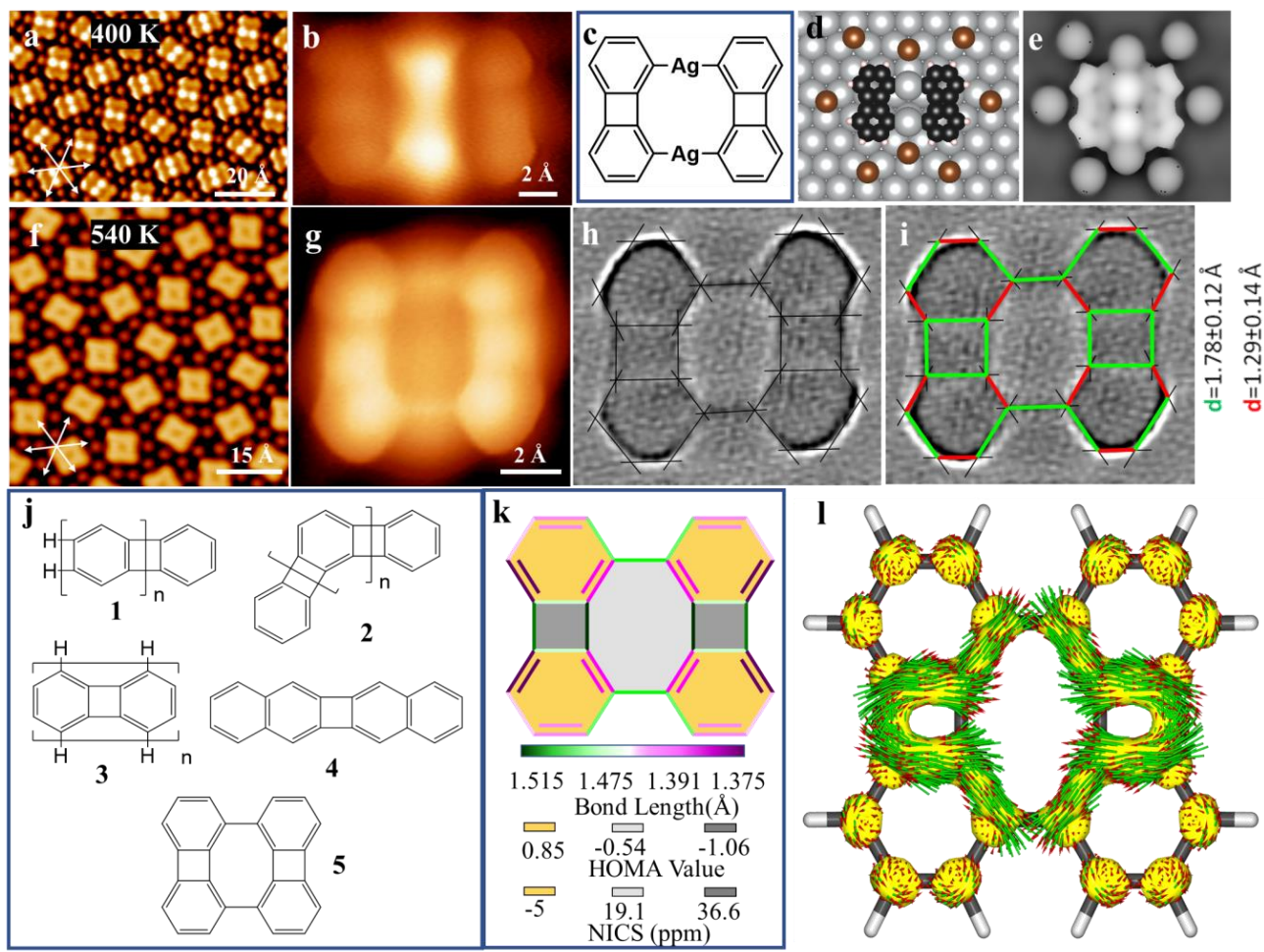

**Figure 2.** (a, b) Overview and constant-height BR-STM images of the organometallic dimer covered sample after annealing the 300 K sample to 400 K. (c, d, e) The structural model, DFT-optimized structure, and simulated STM image of the organometallic dimer, $U_{bias}$ of (e) is set to −0.1 V. (f, g) Overview and constant-height BR-STM images of the sample after further annealing to 540 K. Tunneling parameters: (a) U = −0.1 V, I = 1.5 nA; (b) U = 5 mV. (f) U = 0.1 V, I = 1.2 nA; (g) U = 5 mV. (h, i) Corresponding Laplace filtered image of (g) for a better view of the bonds. The average length for the single bonds (marked in green in panel i) is 1.78±0.12 Å, whereas the average length for the double bonds (marked in red in panel i) is 1.29±0.12 Å. (j) Chemical structures of some biphenylene derivatives showing the bond alternations. (k) DFT calculated C-C bonds lengths, HOMA (on surface) and NICS (in gas phase) values of the biphenylene dimer, represented by gradient colors as indicted. The phenyl rings in the biphenylene dimer have HOMA value of 0.85 and NICS value of -5 ppm, while the eight-membered and four-membered rings have HOMA values of −0.54 and −1.06 and NICS value of 19.1 ppm and 36.6 ppm, respectively. (l) ACID analysis of biphenylene dimer calculated in the gas phase.

Upon annealing the sample at 400 K, large-area close-packed islands composed of organometallic dimers are obtained, as shown in Figure 2a (overview STM image is shown in Figure S4). This is presumably attributed to the planarization of bi-radical biphenylenes on the surface, where two C-Ag-C linkages are formed as expected (Figure 2c).[41] The BR-STM image using a CO functionalized probe[42] corroborates the structure, as shown in Figure 2b. The two bright dots in the middle are assigned to two Ag adatoms, while the darker sides are biphenylenes. In addition, DFT optimized structural model (Figure 2d) and simulated STM image (Figure 2e) both fit the experimental result well. The formation of an organometallic dimer is also supported by the C 1*s* SRPE spectrum which is deconvoluted into C-C and C-Ag components with a ratio of 5.4:1, in good agreement with the proposed molecular structure (ideally 5:1, as derived from the structural model shown in Figure 1c). At these conditions, full debromination of TBBP completes (see Br 3*d* SRPES in Figure 1a). Br adatoms image as relatively dark dots surround the organometallic dimer *via* Br···H hydrogen bonds (Figure 2a).[41]

Further annealing this sample at 540 K triggers the formation of the final covalent product biphenylene dimer (Figure 2f and S4) after the removal of interstitial Ag adatoms from the organometallic dimers. The submolecular structure is clearly characterized by the BR-STM with a CO functionalized probe, as seen in Figure 2g. The covalent connection between two monomers is further confirmed by C 1*s* SRPES where C-Ag signal disappears at 540 K (Figure 1b). Because the annealing from 400 K to 540 K leads to the partial desorption of Br atoms as revealed in the Br 3*d* spectra, the C 1*s* peak shifts back to the high binding energy position at 540 K.[28-29, 32-34]

**Chemical structure of biphenylene dimer.** The biphenylene dimer is of particular interest for studying molecular anti-aromaticity since it is a 4, 6, and 8-membered rings containing structure.[18, 24, 43] A cyclic molecule typically shows antiaromaticity if it holds 4n (n is a positive integer) delocalized electrons while a cyclic molecule containing 4n+2 delocalized electrons is aromatic.[44] In particular, cyclobutadiene is literally thought to be very antiaromatic and not stable. Plenty of theoretical and experimental efforts (mostly crystallography) have been made to study the bond alternation of several biphenylene derivatives.[18, 45-51] A few typical examples are shown in Figure 2j. The linear polyphenylene **1** shows delocalized electronic properties and one double bond can be involved inside the four-membered ring. In contrast, the angular polyphenylene **2** and oligobiphenylenes **3** possess localized π-bonding and hold radialene structures. The double bonds tend to be exocyclic with respect to four-membered rings in **2** and **3** to minimize the antiaromatic character of four-electron π-bonding within the cyclobutadiene.

The first real-space evidence for the bond alternation of biphenylene related structure was given by Kawai *et al.*, by analyzing the bond lengths of molecule **4** using non-contact atomic force microscopy (nc-AFM).[35] Here we demonstrate that the bond alternation of biphenylene dimer **5** is similar to that of biphenylene monomer, that is, the radialene structure is energetically favorable. A Laplace filtered BR-STM image is presented in Figure 2h which enhances the bond feature.[52] We draw the best fitting straight lines along the different bonds of the molecular structure and take the crossing points as reference for the bond length analysis. Remarkably large differences in the bond lengths can be easily identified, that is, those predicted to display a double bond character are clearly shorter. A detailed analysis (Figure 2i) reveals that the average length for the single bonds (marked in green) is 1.78±0.12 Å, whereas the average length for the double bonds (marked in red) is 1.29±0.12 Å, clearly out of the error range of one another. It must be reminded, that these values are not the actual bond lengths. However, the artifact that causes these lengths to deviate from the real values is strongly bond-order dependent, which thus allows for an easy discrimination of the single and double bonds by this comparative analysis.[53] We also calculated the length of each C-C bond of the biphenylene dimer adsorbed on the Ag(111) surface and derived the harmonic oscillator model of aromaticity (HOMA) values (Figure 2k.[54] A higher HOMA value is generally associated with a higher degree of π-electron delocalization and increased aromatic stabilization. The phenyl rings have the highest HOMA value of 0.85; The eight-membered and four-membered rings have HOMA values of only –0.54 and –1.06, respectively, indicating their low aromaticities. This is presumably because of the single bonds involved there, in which decreases the π-electron delocalization on them. Nevertheless, the HOMA value of the four-membered ring is still much higher than the value of cyclobutadiene (–4.277) which indicates the anti-aromaticity in the four-membered ring is significantly reduced in the biphenylene dimer.[54] This is further supported by the nuclear-independent chemical shift (NICS) analysis,[55] as shown in Figure 2k. The phenyl rings are rather *quasi*-nonaromatic (-5 ppm), while the four- and eight-membered rings are highly antiaromatic (36.6 and 19.1 ppm, respectively). In addition, we have investigated the induced currents that result from an applied magnetic field by the anisotropy of the induced current density (ACID).[56] The plot depicted in Figure 2l clearly shows a high anticlockwise current flowing along the central rings including both four- and eight-membered ring, which suggests the anti-aromaticity of the four- and eight-membered ring. In contrast, only fragmented clockwise ring current (weak aromaticity) is shown in phenyl rings, in agreement with their low NICS values. The reduced aromaticity of the phenyl ring is due to the fixed localization of the double bonds caused by the four membered rings.

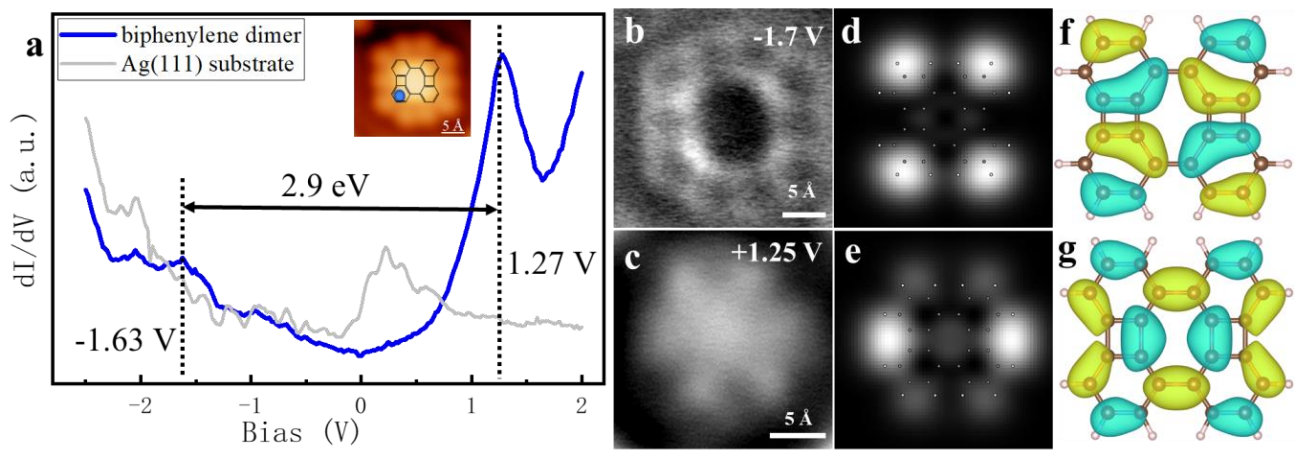

**Figure 3.** Electronic properties of biphenylene dimer on Ag(111). (a) Differential conductance (dI/dV) spectra. The blue curve shows the dI/dV spectrum recorded at the blue dot marked position on the molecule, with a Br functionalized tip. The grey curve shows the reference dI/dV spectrum measured on the bare Ag(111) surface with the same tip. The ten dots surrounding the molecule are ten bromine adatoms. (b, c) Constant-current dI/dV maps taken at the bias voltages of -1.7 and 1.25 V, respectively. Lock-in amplitude, 20 mV; oscillation, 731 Hz; current, 500 pA. (d, e) DFT calculated dI/dV maps corresponded to HOMO and LUMO of biphenylene dimer in gas phase, respectively. (f, g) DFT calculated LDOS distribution of HOMO and LUMO orbitals of biphenylene dimer in gas phase, respectively.

**Electronic structure of biphenylene dimer.** The electronic properties of the biphenylene dimer adsorbed on Ag(111) have been probed by STS, as shown in Figure 3a. The two peaks at -1.63 and 1.27 V should be attributed to the highest occupied and the lowest unoccupied molecular orbitals (HOMO and LUMO), respectively. This is supported by the good agreement between experimentally obtained (Figure 3b and c) and DFT simulated dI/dV maps of the HOMO and LUMO (Figure 3d and e) of biphenylene dimer (the ten dark rings in Figure 3b are from the contribution of the surrounding Br adatoms). Thus, the band gap of the biphenylene dimer adsorbed on Ag(111) is 2.9 eV, which is larger than that of biphenylene ribbon, as reported by Fan et al.,[24] due to its smaller size. The relatively large bandgap is presumably attributed to the relatively large electronic localization on the phenyl groups (weak electronic conjugation between them). The calculated electronic local density of states (LDOS) distributions related to the HOMO and LUMO orbitals of the biphenylene dimer are presented in Figure 3f and 3g. Accordingly, the HOMO is largely localized in the phenyl groups, while the LUMO is mostly distributed in the four-membered ring and the single bond between two biphenylene monomers, thus fitting the proposed bond alternation well, as shown in Figure 2k. The DFT calculated charge density of HOMO and LUMO of biphenylene dimer adsorbing on Ag(111) is presented in Figure S10, very similar to those in gas phase as shown in Figure 3f-g, implying a weak interaction between biphenylene dimer and the Ag(111) surface.

**Reaction pathway analysis.** Next, we focus on the reaction mechanisms of TBBP and DBBP on Ag(111). The optimized structures from a DFT-based model for each reaction step of TBBP are shown in Figure 4a. The details of these calculations are given in the SI.

As shown in Figure 4a, the reaction of TBBP is an exothermic reaction where the energy diagram goes downhill in each reaction step. Notably, the first step is an adsorption-determined spontaneous process, which is also reflected by the considerable heat release of 3.96 eV. After full debromination of TBBP, the intramolecular annulation reaction of the four-radical biphenyl turns out to be the most thermodynamically favorable pathway. This is supported by the fact that the energy of Ag-biphenyl complex, as another possible structure, is 3.75 eV higher than the biphenylene monomer product on Ag(111), as seen in Figure S15. In fact, the energy barrier of the intramolecular annulation of TBBP on Ag(111) should be from the debromination reaction of TBBP. As shown in Figure 4b, two Br atoms at the same benzene ring (either 1 and 2, or 3 and 4 sites) prefer dissociating simultaneously with a transition energy barrier of 0.72 eV and heat releasing of 1.25 eV. This excludes the possibility of the intramolecular [2+2] annulation may also occur as long as two bromines at the same side (either 1 and 3, or 2 and 4 sites) of molecular backbone dissociated (instead of full debromination).

The following steps are the conventional planarization of bi-radical biphenylene monomer with the help of Ag adatoms and the subsequent Ullmann coupling by thermal treatments, finally forming biphenylene dimer.

In contrast, for the reaction of DBBP on Ag(111), the reaction is an Ullmann coupling followed by the cyclodehydrogenation, forming dibenzo[e,l]pyrene (Scheme 1). It is obvious that the difference between the two reactions (TBBP and DBBP) on Ag(111) originates from different molecular adsorption configurations. Different from the spontaneous annulation of TBBP, in case of DBBP, both the Ag surface and Ag adatoms are adequate to stabilize bi-radical biphenyl, resulting in the formation of the organometallic dimer with four-fold C-Ag bonds (Figure S18). It is known that Ullmann coupling has a very low reaction barrier after the removal of interstitial Ag atoms from organometallic intermediate,[57-59] thus the formation of dibenzo[e,l]pyrene is expected. Each reaction step of DBBP on Ag(111) is an exothermic reaction, as seen in Figure S18.

It is worth noting that the intramolecular annulation reaction of bi-radical biphenyl should be also possible after removal of Ag atoms from the organometallic dimer. However, a much higher temperature is needed for the intramolecular annulation of bi-radical biphenyl. For example, in the work of Kawai *et al.*, intramolecular annulation to form **4** in Figure 2j could be achieved only at temperatures higher than 406 K.[35] This should be higher than that of the Ullmann coupling between radicals (usually < 1 eV; demetallization is normally the rate- determining step of Ullmann reaction on surfaces).[58-59] The energy barrier of intramolecular annulation of bi-radical biphenyl relates to the planarization of the molecule on the Ag(111) surface. The relatively high barrier of intramolecular annulation of DBBP on Ag(111) is also supported by the experimental results. First of all, annealing of DBBP on Ag(111) at 450 K leading to the formation of only dibenzo[e,l]pyrene and no biphenylene products were observed. However, one could argue that the self-assembly template effect of an organometallic dimer island might enhance the Ullmann coupling which takes place through a molecular transition and prohibits the intramolecular annulation which occurs by molecular rotation.[5, 60] Therefore, we did control experiments by depositing DBBP molecules on the hot Ag(111) surfaces to avoid self-assembly template effects. The results show that the products are still almost 100% dibenzo[e,l]pyrene by 450 K hot deposition (Figure S5). The biphenylene monomer from intramolecular annulation can appear as a very minor byproduct only when the deposition was employed at a hotter sample (550 K), together with some other byproducts (Figure S6). This is because the deposition on a very hot sample can normally overcome some high reaction barriers.[6, 61-62] Nevertheless, the total yield of these byproducts is less than 10% and dibenzo[e,l]pyrene still dominates, thus supporting the point that the energy barrier of the Ullmann coupling of DBBP on Ag(111) is much lower as compared to that of intramolecular annulation.

To investigate the generality of the chemisorption induced reaction selectivity of TBBP on metal surfaces, we further studied the reaction of TBBP on the Ag(100) surface.

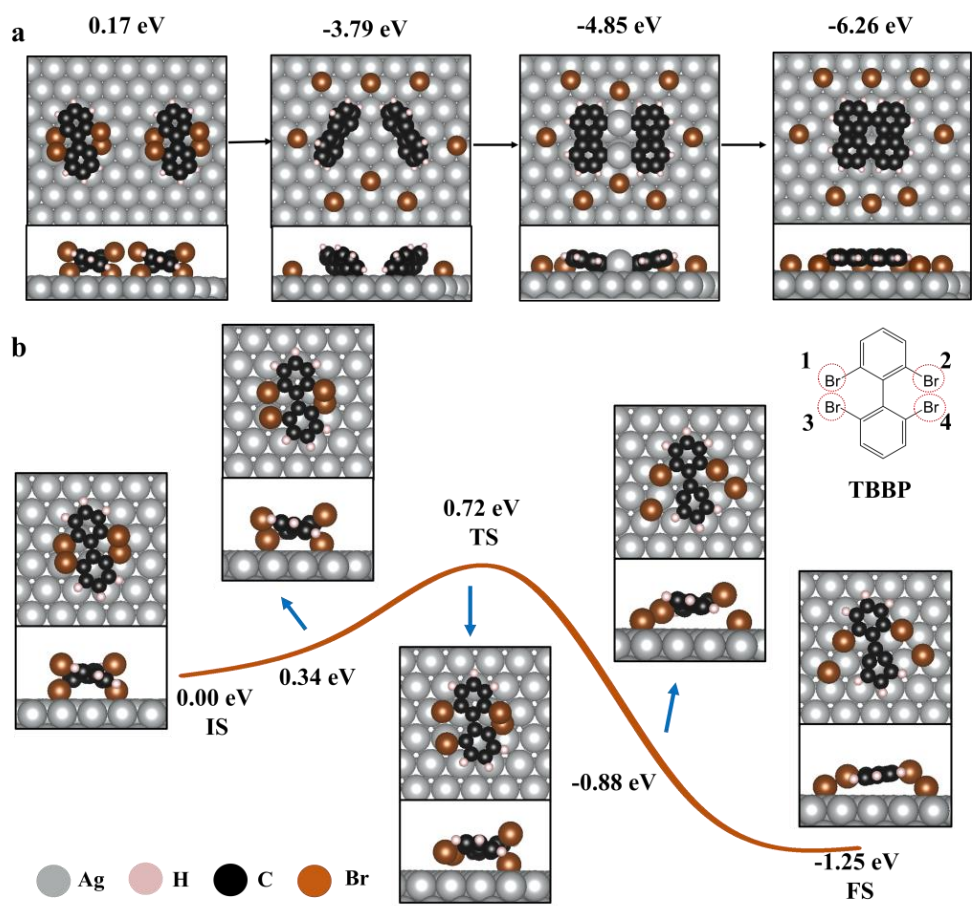

**Figure 4.** (a) Optimized configurations of the reaction evolution of TBBP to biphenylene dimer on Ag(111). The energy values shown in each structure are calculated with the Eq S3 as shown in SI, where the total energy of a TBBP monomer adsorbate is taken as a reference for the comparison between different adsorbates. (b) Energy diagrams of the initial debromination reaction of TBBP on Ag(111). The initial state, transition state and final state are denoted as IS, TS and FS, respectively. The structure model of TBBP is shown in the upper right corner in which four bromine atoms are marked by dashed red circles. The total energy of initial state is taken as a reference (set as 0). Color code: C, black; Ag, grey; H, pink; Br, brown.

Biphenylene dimers are also selectively obtained on Ag(100), by undergoing an organometallic intermediate state, identical as the reaction pathway of TBBP on Ag(111), as shown in Figure S7.

In short, the main reason for the different reaction pathways of TBBP and DBBP on Ag(111) is from their different chemisorption configurations. Four-radical biphenyl takes a spontaneous intramolecular annulation to lower the overall energy. In sharp contrast, bi-radical biphenyl can be easily stabilized by the surface adatoms. In addition, intramolecular annulation of DBBP is less favored than the competing Ullmann coupling, thus leading to its completely different reaction selectivity on Ag(111) as compared to TBBP.

**Conclusions**

In summary, biphenylene dimers are selectively synthesized on a Ag(111) surface with a high yield, starting from TBBP. Using BR-STM and STS, we demonstrate that the radialene rather than a cyclobutadiene structure is preferred in the biphenylene dimer. The high selectivity toward a biphenylene dimer is attributed to the special adsorption configuration of TBBP on Ag(111). After debromination, the four-radical biphenyl cannot be stabilized simply with the help of surface atoms, which instead undergoes intramolecular annulation reaction and finally forms a biphenylene dimer *via* intermolecular Ullmann coupling. In contrast, the bi-radical biphenyl from debrominations of DBBP can be efficiently stabilized by surface Ag adatoms, forming an organometallic dimer. This organometallic intermediate subsequently reacts into dibenzo[e,l]pyrene *via* Ullmann coupling. Control experiments demonstrate that the energy barrier of intramolecular annulation reaction is higher than that of the Ullmann coupling reaction for bi-radical biphenyl on Ag(111). The different adsorption configurations on Ag(111) lead to different reaction pathways for the two structurally similar adsorbates. We believe this work can invoke scientists' interests since it serves as an example on how one can control the reaction selectivity for given adsorbate by tuning its adsorption behavior.

More interestingly, based on STS, and the combined HOMA, NICS, and ACIDS analyses, we provide comprehensive interpretations toward the antiaromaticities of four- and eight- membered rings, which contain 4n electrons. In addition, because of the bond-confinement caused by the existence of four-membered ring, the aromaticity of phenyl is significantly reduced. This bond-confinement effect revealed here could be potentially employed for other graphene-based and non-benzenoid carbon structures, tuning the electronic properties and chemical reactivity of these materials.

## ASSOCIATED CONTENT

**Supporting Information**. Detailed descriptions of experimental and theoretical methods, additional STM images. This material is available free of charge via the Internet at http://pubs.acs.org.

## AUTHOR INFORMATION

Corresponding Author

* J. F. Zhu, jfzhu@ustc.edu.cn

* J.-S. McEwen, js.mcewen@wsu.edu

* T. Wang, taowang@dipc.org

Author Contributions

‡These authors contributed equally. The manuscript was written through contributions of all authors. All authors have given approval to the final version of the manuscript.

Notes

The authors declare no competing financial interest.

## ACKNOWLEDGMENT

This work was financially supported by the National Natural Science Foundation of China (21773222, 51772285, U1732272, and U1932214), the National Key R&D Program of China (2017YFA0403402, 2017YFA0403403, and 2019YFA0405601), Users with Excellence Program of Hefei Science Center CAS (2020HSC-UE004). The work at Washington State University was primarily funded through that National Science Foundation CAREER program under contract number CBET-1653561. This work was also partially funded by the Joint Center for Deployment and Research in Earth Abundant Materials (JCDREAM) in Washington State. Most of the computational resources was provided by the Kamiak HPC under the Center for Institutional Research Computing at Washington State University. A portion of the computer time for the computational work was performed using EMSL, a national scientific user facility sponsored by the Department of Energy's Office of Biological and Environmental Research and located at Pacific Northwest National Laboratory. This research also used resources of the National Energy Research Scientific Computing Center (NERSC), a U.S. Department of Energy Office of Science User Facility operated under Contract No. DE-AC02-05CH11231. The work at DIPC was primarily funded through Juan de la Cierva grant (No. FJC2019-041202-I) from Spanish Ministry of Economy and Competitiveness, the European Union's Horizon 2020 research and innovation program (Marie Skłodowska-Curie Actions Individual Fellowship (No. 101022150).

## REFERENCES

(1) Clair, S.; de Oteyza, D. G. Controlling a Chemical Coupling Reaction on a Surface: Tools and Strategies for On-Surface Synthesis. *Chem. Rev.* **2019**, *119*, 4717-4776.

(2) Wang, T.; Zhu, J. F. Confined On-Surface Organic Synthesis: Strategies and Mechanisms. *Surf. Sci. Rep.* **2019**, *74*, 97-140.

(3) Han, D.; Zhu, J. F. Surface-Assisted Fabrication of Low-Dimensional Carbon-Based Nanoarchitectures. *J. Phys.: Condens. Matter* **2021**, *33*, 343001.

(4) Li, Q.; Gao, J.; Li, Y.; Fuentes-Cabrera, M.; Liu, M.; Qiu, X.; Lin, H.; Chi, L.; Pan, M. Self-Assembly Directed One-Step Synthesis of [4] Radialene on Cu(100) Surfaces. *Nat. Commun.* **2018**, *9*, 3113.

(5) Chen, Q.; Cramer, J. R.; Liu, J.; Jin, X.; Liao, P.; Shao, X.; Gothelf, K. V.; Wu, K. Steering On-Surface Reactions by a Self-Assembly Approach. *Angew. Chem., Int. Ed.* **2017**, *56*, 5026-5030.

(6) Wang, T.; Huang, J.; Lv, H.; Fan, Q.; Feng, L.; Tao, Z.; Ju, H.; Wu, X.; Tait, S. L.; Zhu, J. Kinetic Strategies for the Formation of Graphyne Nanowires via Sonogashira Coupling on Ag(111). *J. Am. Chem. Soc.* **2018**, *140*, 13421-13428.

(7) Liu, W.; Luo, X.; Bao, Y.; Liu, Y. P.; Ning, G. H.; Abdelwahab, I.; Li, L.; Nai, C. T.; Hu, Z. G.; Zhao, D.; Liu, B.; Quek, S. Y.; Loh, K. P. A Two-Dimensional Conjugated Aromatic Polymer via C-C Coupling Reaction. *Nat. Chem.* **2017**, *9*, 563-570.

(8) Sánchez-Sánchez, C.; Dienel, T.; Deniz, O.; Ruffieux, P.; Berger, R.; Feng, X.; Müllen, K.; Fasel, R. Purely Armchair or Partially Chiral: Noncontact Atomic Force Microscopy Characterization of Dibromo-Bianthryl-Based Graphene Nanoribbons Grown on Cu(111). *ACS Nano* **2016**, *10*, 8006-8011.

(9) Kong, H.; Yang, S.; Gao, H.; Timmer, A.; Hill, J. P.; Arado, O. D.; Monig, H.; Huang, X.; Tang, Q.; Ji, Q.; Liu, W.; Fuchs, H. Substrate-Mediated C-C and C-H Coupling after Dehalogenation. *J. Am. Chem. Soc.* **2017**, *139*,

3669-3675.
(10) Shi, K. J.; Shu, C. H.; Wang, C. X.; Wu, X. Y.; Tian, H.; Liu, P. N. On-Surface Heck Reaction of Aryl Bromides with Alkene on Au(111) with Palladium as Catalyst. *Org. Lett.* **2017**, *19*, 2801-2804.
(11) Moreno, C.; Panighel, M.; Vilas-Varela, M.; Sauthier, G.; Tenorio, M.; Ceballos, G.; Peña, D.; Mugarza, A. Critical Role of Phenyl Substitution and Catalytic Substrate in the Surface-Assisted Polymerization of Dibromobianthracene Derivatives. *Chem. Mater.* **2019**, *31*, 331-341.
(12) Merino-Díez, N.; Pérez Paz, A.; Li, J.; Vilas-Varela, M.; Lawrence, J.; Mohammed, M. S. G.; Berdonces-Layunta, A.; Barragán, A.; Pascual, J. I.; Lobo-Checa, J.; Peña, D.; de Oteyza, D. G. Hierarchy in the Halogen Activation During Surface-Promoted Ullmann Coupling. *ChemPhysChem* **2019**, *20*, 2305-2310.
(13) Zhong, Q.; Ebeling, D.; Tschakert, J.; Gao, Y.; Bao, D.; Du, S.; Li, C.; Chi, L.; Schirmeisen, A. Symmetry Breakdown of 4,4"-Diamino-p-Terphenyl on a Cu(111) Surface by Lattice Mismatch. *Nat. Commun.* **2018**, *9*, 3277.
(14) Grill, L.; Dyer, M.; Lafferentz, L.; Persson, M.; Peters, M. V.; Hecht, S. Nano-Architectures by Covalent Assembly of Molecular Building Blocks. *Nat. Nanotechnol.* **2007**, *2*, 687-691.
(15) Cai, J.; Ruffieux, P.; Jaafar, R.; Bieri, M.; Braun, T.; Blankenburg, S.; Muoth, M.; Seitsonen, A. P.; Saleh, M.; Feng, X.; Müllen, K.; Fasel, R. Atomically Precise Bottom-Up Fabrication of Graphene Nanoribbons. *Nature* **2010**, *466*, 470-473.
(16) Zhou, X.; Wang, C.; Zhang, Y.; Cheng, F.; He, Y.; Shen, Q.; Shang, J.; Shao, X.; Ji, W.; Chen, W.; Xu, G.; Wu, K. Steering Surface Reaction Dynamics with a Self-Assembly Strategy: Ullmann Coupling on Metal Surfaces. *Angew. Chem., Int. Ed.* **2017**, *56*, 12852-12856.
(17) Feng, L.; Wang, T.; Jia, H.; Huang, J.; Han, D.; Zhang, W.; Ding, H.; Xu, Q.; Du, P.; Zhu, J. On-Surface Synthesis of Planar Acenes via Regioselective Aryl-Aryl Coupling. *Chem. Commun.* **2020**, *56*, 4890-4893.
(18) Rajca, A.; Safronov, A.; Rajca, S.; Ross, C. R.; Stezowski, J. J. Biphenylene Dimer. Molecular Fragment of a Two-Dimensional Carbon Net and Double-Stranded Polymer. *J. Am. Chem. Soc.* **1996**, *118*, 7272-7279.
(19) Zhang, C.; Kazuma, E.; Kim, Y. Atomic-Scale Visualization of the Stepwise Metal-Mediated Dehalogenative Cycloaddition Reaction Pathways: Competition between Radicals and Organometallic Intermediates. *Angew. Chem., Int. Ed.* **2019**, *58*, 17736-17744.
(20) Tran, B. V.; Pham, T. A.; Grunst, M.; Kivala, M.; Stohr, M. Surface-confined [2 + 2] cycloaddition towards one-dimensional polymers featuring cyclobutadiene units. *Nanoscale* **2017**, *9*, 18305-18310.
(21) Li, D. Y.; Qiu, X.; Li, S. W.; Ren, Y. T.; Zhu, Y. C.; Shu, C. H.; Hou, X. Y.; Liu, M.; Shi, X. Q.; Qiu, X.; Liu, P. N. Ladder Phenylenes Synthesized on Au(111) Surface via Selective [2+2] Cycloaddition. *J. Am. Chem. Soc.* **2021**, *143*, 12955-12960.
(22) Liu, M.; Liu, M.; She, L.; Zha, Z.; Pan, J.; Li, S.; Li, T.; He, Y.; Cai, Z.; Wang, J.; Zheng, Y.; Qiu, X.; Zhong, D. Graphene-like Nanoribbons Periodically Embedded with Four- and Eight-Membered Rings. *Nat Commun.* **2017**, *8*, 14924.
(23) Zhang, R.; Xia, B.; Xu, H.; Lin, N. Identifying Multinuclear Organometallic Intermediates in On-Surface [2+2] Cycloaddition Reactions. *Angew. Chem., Int. Ed.* **2019**, *58*, 16485-16489.
(24) Fan, Q.; Yan, L.; Tripp, M. W.; Krejci, O.; Dimosthenous, S.; Kachel, S. R.; Chen, M.; Foster, A. S.; Koert, U.; Liljeroth, P.; Gottfried, J. M. Biphenylene network: A nonbenzenoid carbon allotrope. *Science* **2021**, *372*, 852-856.
(25) Sanchez-Sanchez, C.; Nicolai, A.; Rossel, F.; Cai, J.; Liu, J.; Feng, X.; Mullen, K.; Ruffieux, P.; Fasel, R.; Meunier, V. On-Surface Cyclization of ortho-Dihalotetracenes to Four- and Six-Membered Rings. *J. Am. Chem. Soc.* **2017**, *139*, 17617-17623.
(26) Sanchez-Sanchez, C.; Dienel, T.; Nicolai, A.; Kharche, N.; Liang, L.; Daniels, C.; Meunier, V.; Liu, J.; Feng, X.; Mullen, K.; Sanchez-Valencia, J. R.; Groning, O.; Ruffieux, P.; Fasel, R. On-Surface Synthesis and Characterization of Acene-Based Nanoribbons Incorporating Four-Membered Rings. *Chemistry* **2019**, *25*, 12074-12082.
(27) de Oteyza, D. G.; Garcia-Lekue, A.; Vilas-Varela, M.; Merino-Diez, N.; Carbonell-Sanroma, E.; Corso, M.; Vasseur, G.; Rogero, C.; Guitian, E.; Pascual, J. I.; Ortega, J. E.; Wakayama, Y.; Peña, D. Substrate-Independent Growth of Atomically Precise Chiral Graphene Nanoribbons. *ACS Nano* **2016**, *10*, 9000-9008.
(28) Wang, T.; Lv, H.; Huang, J.; Shan, H.; Feng, L.; Mao, Y.; Wang, J.; Zhang, W.; Han, D.; Xu, Q.; Du, P.; Zhao, A.; Wu, X.; Tait, S. L.; Zhu, J. Reaction Selectivity of Homochiral versus Heterochiral Intermolecular Reactions of Prochiral Terminal Alkynes on Surfaces. *Nat. Commun.* **2019**, *10*, 4122.
(29) Fan, Q.; Liu, L.; Dai, J.; Wang, T.; Ju, H.; Zhao, J.; Kuttner, J.; Hilt, G.; Gottfried, J. M.; Zhu, J. Surface Adatom Mediated Structural Transformation in Bromoarene Monolayers: Precursor Phases in Surface Ullmann Reaction. *ACS Nano* **2018**, *12*, 2267-2274.
(30) Simonov, K. A.; Generalov, A. V.; Vinogradov, A. S.; Svirskiy, G. I.; Cafolla, A. A.; McGuinness, C.; Taketsugu, T.; Lyalin, A.; Martensson, N.; Preobrajenski, A. B. Synthesis of Armchair Graphene Nanoribbons from The 10,10'-dibromo-9,9'-bianthracene Molecules on Ag(111): The Role of Organometallic Intermediates. *Surf. Sci. Rep.* **2018**, *8*, 3506.
(31) Abyazisani, M.; MacLeod, J. M.; Lipton-Duffin, J. Cleaning up after the Party: Removing the Byproducts of On-Surface Ullmann Coupling. *ACS Nano* **2019**, *13*, 9270-9278.
(32) Fan, Q. T.; Wang, C. C.; Liu, L. M.; Han, Y.; Zhao, J.; Zhu, J. F.; Kuttner, J.; Hilt, G.; Gottfried, J. M. Covalent, Organometallic, and Halogen-Bonded Nanomeshes from Tetrabromo-Terphenyl by Surface-Assisted Synthesis on Cu(111). *J. Phys. Chem. C* **2014**, *118*, 13018-13025.
(33) Wang, T.; Lv, H. F.; Feng, L.; Tao, Z. J.; Huang, J. M.; Fan, Q. T.; Wu, X. J.; Zhu, J. F. Unravelling the Mechanism of Glaser Coupling Reaction on Ag(111) and Cu(111) Surfaces: a Case for Halogen Substituted Terminal Alkyne. *J. Phys. Chem. C* **2018**, *122*, 14537-14545.
(34) Han, D.; Fan, Q. T.; Dai, J. Y.; Wang, T.; Huang, J. M.; Xu, Q.; Ding, H. H.; Hu, J.; Feng, L.; Zhang, W. Z.; Zeng, Z. W.; Gottfried, J. M.; Zhu, J. F. On-Surface Synthesis of Armchair-Edged Graphene Nanoribbons with Zigzag Topology. *J. Phys. Chem. C* **2020**, *124*, 5248-5256.
(35) Kawai, S.; Takahashi, K.; Ito, S.; Pawlak, R.; Meier, T.; Spijker, P.; Canova, F. F.; Tracey, J.; Nozaki, K.; Foster, A. S.; Meyer, E. Competing Annulene and Radialene Structures in a Single Anti-Aromatic Molecule Studied by High-Resolution Atomic Force Microscopy. *ACS Nano* **2017**, *11*, 8122-8130.
(36) Hieulle, J.; Carbonell-Sanromà, E.; Vilas-Varela, M.; Garcia-Lekue, A.; Guitián, E.; Peña, D.; Pascual, J. I. On-Surface Route for Producing Planar Nanographenes with Azulene Moieties. *Nano Lett.* **2018**, *18*, 418-423.
(37) Wang, T.; Fan, Q.; Feng, L.; Tao, Z.; Huang, J.; Ju, H.; Xu, Q.; Hu, S.; Zhu, J. Chiral Kagome Lattices from On-Surface Synthesized Molecules. *ChemPhysChem* **2017**, *18*, 3329-3333.
(38) Wang, T.; Lawrence, J.; Sumi, N.; Robles, R.; Castro-Esteban, J.; Rey, D.; Mohammed, M. S. G.; Berdonces-Layunta, A.; Lorente, N.; Pérez, D.; Peña, D.; Corso, M.; de Oteyza, D. G. Challenges in the Synthesis of Corannulene-Based Non-Planar Nanographenes on Au(111) Surfaces. *Phys. Chem. Chem. Phys.* **2021**, *23*, 10845-10851.
(39) Beniwal, S.; Chen, S.; Kunkel, D. A.; Hooper, J.; Simpson, S.; Zurek, E.; Zeng, X. C.; Enders, A. Kagome-like Lattice of pi-pi Stacked 3-Hydroxyphenalenone on Cu(111). *Chem. Commun.* **2014**, *50*, 8659-62.
(40) Inayeh, A.; Groome, R. R. K.; Singh, I.; Veinot, A. J.; de Lima, F. C.; Miwa, R. H.; Crudden, C. M.; McLean, A. B. Self-Assembly of N-Heterocyclic Carbenes on Au(111). *Nat Commun.* **2021**, *12*, 4034.
(41) Fan, Q.; Gottfried, J. M.; Zhu, J. Surface-Catalyzed C-C Covalent Coupling Strategies toward the Synthesis of Low-Dimensional Carbon-Based Nanostructures. *Acc. Chem. Res.* **2015**, *48*, 2484-2494.
(42) Jelínek, P. High Resolution SPM Imaging of Organic Molecules With Functionalized Tips. *J. Phys.: Condens. Matter* **2017**, *29*, 343002-343019.
(43) Randić, M.; Balaban, A. T.; Plavšić, D. Applying the Conjugated Circuits Method to Clar Structures of [n] Phenylenes for Determining Resonance Energies. *Phys. Chem. Chem. Phys.* **2011**, *13*, 20644-20648.
(44) Rosenberg, M.; Dahlstrand, C.; Kilsa, K.; Ottosson, H. Excited State Aromaticity and Antiaromaticity: Opportunities for Photophysical and Photochemical Rationalizations. *Chem. Rev.* **2014**, *114*, 5379-5425.
(45) Barron, T. H. K.; Barton, J. W.; Johnson, J. D. On the stability of some polycyclic biphenylene derivatives. *Tetrahedron* **1966**, *22*, 2609-2613.
(46) Yokozeki, A.; Wilcox, C. F.; Bauer, S. H. Biphenylene. Internuclear Distances and Their Root Mean Square Amplitudes of Vibration. *J. Am. Chem. Soc.* **1974**, *96*, 1026-1032.
(47) Schmidt-Radde, R. H.; Vollhardt, K. P. C. The Total Synthesis of Angular [4]- and [5] Phenylene. *J. Am. Chem. Soc.* **1992**, *114*, 9713-9715.
(48) Baldridge, K. K.; Siegel, J. S. Bond Alternation in Triannelated Benzenes: Dissection of Cyclic .Pi. from Mills-Nixon Effects. *J. Am. Chem. Soc.* **1992**, *114*, 9583-9587.
(49) Faust, R.; Glendening, E. D.; Streitwieser, A.; Vollhardt, K. P. C. Ab Initio Study of .Sigma.- and .Pi.-Effects in Benzenes Fused to Four-Membered Rings: Rehybridization, Delocalization, and Antiaromaticity. *J. Am. Chem. Soc.* **1992**, *114*, 8263-8268.
(50) Bürgi, H.-B.; Baldridge, K. K.; Hardcastle, K.; Frank, N. L.; Gantzel, P.; Siegel, J. S.; Ziller, J. X-Ray Diffraction Evidence for a Cyclohexatriene Motif in the Molecular Structure of Tris(bicyclo[2.1.1]hexeno)benzene:

Bond Alternation after the Refutation of the Mills–Nixon Theory. *Angew. Chem., Int. Ed.* **1995**, *34*, 1454-1456.

(51) Vollhardt, K. P. C. The Phenylenes. *Pure Appl. Chem.* **1993**, *65*, 153-156.

(52) Gross, L.; Mohn, F.; Moll, N.; Schuler, B.; Criado, A.; Guitian, E.; Pena, D.; Gourdon, A.; Meyer, G. Bond-Order Discrimination by Atomic Force Microscopy. *Science* **2012**, *337*, 1326-9.

(53) Pozo, I.; Majzik, Z.; Pavlicek, N.; Melle-Franco, M.; Guitian, E.; Pena, D.; Gross, L.; Perez, D. Revisiting Kekulene: Synthesis and Single-Molecule Imaging. *J. Am. Chem. Soc.* **2019**, *141*, 15488-15493.

(54) Setiawan, D.; Kraka, E.; Cremer, D. Quantitative Assessment of Aromaticity and Antiaromaticity Utilizing Vibrational Spectroscopy. *J. Org. Chem.* **2016**, *81*, 9669-9686.

(55) Schleyer, P. V. R.; Maerker, C.; Dransfeld, A.; Jiao, H.; van Eikema Hommes, N. J. R. Nucleus-Independent Chemical Shifts: A Simple and Efficient Aromaticity Probe. *J. Am. Chem. Soc.* **1996**, *118*, 6317-6318.

(56) Herges, R.; Geuenich, D. Delocalization of Electrons in Molecules. *J. Phys. Chem. A* **2001**, *105*, 3214-3220.

(57) Björk, J. Reaction Mechanisms for On-Surface Synthesis of Covalent Nanostructures. *J. Phys.: Condens. Matter* **2016**, *28*, 083002.

(58) Di Giovannantonio, M.; Tomellini, M.; Lipton-Duffin, J.; Galeotti, G.; Ebrahimi, M.; Cossaro, A.; Verdini, A.; Kharche, N.; Meunier, V.; Vasseur, G.; Fagot-Revurat, Y.; Perepichka, D. F.; Rosei, F.; Contini, G. Mechanistic Picture and Kinetic Analysis of Surface-Confined Ullmann Polymerization. *J. Am. Chem. Soc.* **2016**, *138*, 16696-16702.

(59) Björk, J.; Hanke, F.; Stafstrom, S. Mechanisms of Halogen-Based Covalent Self-Assembly on Metal Surfaces. *J. Am. Chem. Soc.* **2013**, *135*, 5768-5575.

(60) Huang, J.; Pan, Y.; Wang, T.; Cui, S.; Feng, L.; Han, D.; Zhang, W.; Zeng, Z.; Li, X.; Du, P.; Wu, X.; Zhu, J. Topology Selectivity in On-Surface Dehydrogenative Coupling Reaction: Dendritic Structure versus Porous Graphene Nanoribbon. *ACS Nano* **2021**, *15*, 4617-4626.

(61) Cirera, B.; Giménez-Agulló, N.; Björk, J.; Martínez-Peña, F.; Martin-Jimenez, A.; Rodriguez-Fernandez, J.; Pizarro, A. M.; Otero, R.; Gallego, J. M.; Ballester, P.; Galan-Mascaros, J. R.; Ecija, D. Thermal Selectivity of Intermolecular versus Intramolecular Reactions on Surfaces. *Nat. Commun.* **2016**, *7*, 11002.

(62) Lin, T.; Zhang, L.; Björk, J.; Chen, Z.; Ruben, M.; Barth, J. V.; Klappenberger, F. Terminal Alkyne Coupling on a Corrugated Noble Metal Surface: From Controlled Precursor Alignment to Selective Reactions. *Chem. Eur. J.* **2017**, *23*, 15588-15593.

### TABLE OF CONTENTS/ABSTRACT GRAPHIC

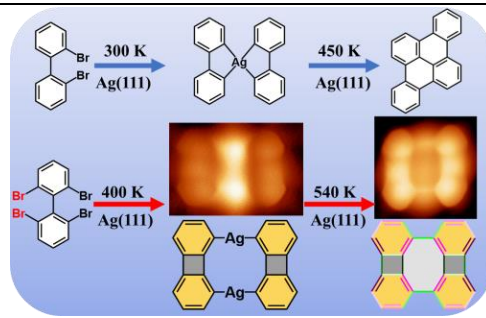